\documentclass[prb,preprint]{revtex4-1} % style for Physical Review B and AJP are similar

\usepackage{amsmath} % need for subequations

\usepackage{graphicx} % for figures

\usepackage{textcomp}

\begin{document}

\title{A Possible Intuitive Derivation of the Kerr Metric in Orthogonal Form Based On Ellipsoidal Metric Ansatz}

\author{Branislav D. Nikoli\'c}
\email{branislav.nikol@gmail.com}
\affiliation{Bonn International Graduate School in Physics and Astronomy (BIGS-PA),\\
University of Bonn,
Germany}

\author{Milan R. Panti\'c}
\email{mpantic@df.ns.ac.rs}
\affiliation{University of Novi Sad, Department of Physics, Novi Sad,
Serbia}
\date{\today}

\begin{abstract}
In this paper we show that it is possible to derive the Kerr solution in an alternative, intuitive way, based on physical reasoning and starting from an orthogonal metric ansatz having manifest ellipsoidal space-time symmetry (ellipsoidal symmetry). This is possible because both flat metric in oblate spheroidal (ellipsoidal) coordinates and Kerr metric in Boyer-Lindquist coordinates can be rewritten in such a form that the difference between the two is only in the time-time and radial-radial metric tensor components, just as is the case with Schwarzschild metric and flat metric in spherical coordinates.
\end{abstract}

\maketitle

\section{Introduction}
When introduced with the Kerr metric for the first time, students often have problems with understanding how it is derived, because most textbooks on General Relativity do not present the original derivation by Kerr\cite{Kerr63}, which is complicated and unintuitive, or any other derivation. However, students can refer to Chandrasekhar\cite{Chandra83}, but this derivation starts from an axisymmetric metric proposed with symmetry arguments, which introduces five unknown functions to be found by solving the Einstein equations. Also, students may wonder why these derivations are so different from Schwarzschild metric derivation and why there is no ``simple'' derivation. On the other hand, there are elegant examples of derivation which uses physical symmetry and gauge arguments such as the derivation by Deser and Franklin\cite{Deser10}, or the derivation by Enderlein\cite{Ender97} which makes use of the Lorentz-transformed basis of 1-forms for a flat space---time in oblate spheroidal coordinates. One is then tempted and encouraged to try to find an intuitive derivation, backed-up with physical arguments, in order to show to students that a simple physical reasoning and observation of some features of Schwarzschild and Kerr solutions can lead to a pedagogical introduction to Kerr metric, covering the derivation. This paper aims to achieve that and may serve as a guideline for students.
Students are usually introduced with the Kerr metric in Boyer-Lindquist coordinates\cite{BoyerLind68}:
\begin{eqnarray}
\label{eqn:metricBL} 
ds^{2} & = & \left( 1-\frac{2Mr}{\rho^{2}} \right)  dt^{2}+\frac{4Mra \sin^{2} \theta}{\rho^{2}}dtd\phi \nonumber \\ & & -\frac{\rho^{2}}{\Delta}dr^{2}-\rho^{2}d\theta^{2}-\left( r^{2}+a^{2}+\frac{2Mra^{2}\sin^{2} \theta}{\rho^{2}}   \right) \sin^{2} \theta d\phi^{2}
\end{eqnarray}
where $M$ and $a$ are mass and angular momentum per unit mass, respectively, and
\begin{eqnarray}
\rho^{2} & = & r^{2}+a^{2}\cos^{2}\theta  \\
\Delta & = & r^{2}-2Mr+a^{2} 
\end{eqnarray}
The metric tensor in these coordinates has the following components:
\begin{eqnarray}
g_{00} & = & 1-\frac{2Mr}{\rho^{2}}  \nonumber \\
g_{03} & = & \frac{4Mra\sin^{2}\theta}{\rho^{2}} \nonumber \\
g_{11} & = & -\frac{\rho^{2}}{\Delta} \nonumber \\
g_{22} & = & -\rho^{2} \nonumber \\ 
g_{33} & = & -\left( r^{2}+a^{2}+\frac{2Mra^{2}\sin^{2} \theta}{\rho^{2}}   \right) \sin^{2} \theta
\end{eqnarray}
Unlike the original derivation by Kerr\cite{Kerr63} and the derivation by Chandrasekhar\cite{Chandra83}, we would like to derive the Kerr metric in a similar manner as Schwarzschild derived the spherically symmetric solution. Schwarzschild derived his solution by proposing the spherically symmetric ansatz:
\begin{equation}
\label{eqn:flatsphans}
ds^{2}=e^{2\rho(r)}dt^{2}-e^{2\sigma(r)}dr^{2}-r^{2}d\theta^{2}-r^{2}\sin^{2}\theta d\phi^{2}
\end{equation}
from which he obtained what is known as the Schwarzschild solution:
\begin{equation}
\label{eqn:schmet}
ds^{2}=\left( 1-\frac{2M}{r}\right) dt^{2}-\left( 1-\frac{2M}{r}\right)^{-1}dr^{2}-r^{2}d\theta^{2}-r^{2}\sin^{2}\theta d\phi^{2}
\end{equation}
where $M$ is the mass parameter. This means that the symmetry of the Minkowski metric in spherical coordinates:
\begin{equation}
\label{eqn:flatsph}
ds^{2}=dt^{2}-dr^{2}-r^{2}d\theta^{2}-r^{2}\sin^{2}\theta d\phi^{2}
\end{equation}
is preserved for the space-time curved by a body of a mass $M$ obeying spherical symmetry. The only differences between Eq.~(\ref{eqn:schmet}) and Eq.~(\ref{eqn:flatsph}) are in the time-time and radial-radial metric tensor components.

In this paper we will follow the similar logic and ask for the solution to the Einstein equations which manifestly preserves the symmetry of an empty ellipsoidal space-time\cite{Krasin78}:
\begin{equation}
\label{eqn:flatellip} 
ds^{2}=dt^{2}-\frac{\rho^{2}}{r^{2}+a^{2}}dr^{2}-\rho^{2}d\theta^{2}-(r^{2}+a^{2})\sin^{2}\theta d\phi^{2}
\end{equation}
On the other hand, we know that the Kerr solution possesses axial symmetry, which can be seen from Eq.~(\ref{eqn:metricBL}), and unlike the Schwarzschild solution given by Eq.~(\ref{eqn:schmet}), the Kerr metric~(\ref{eqn:metricBL}) has the cross term $dtd\phi$, which is not present in Eq.~(\ref{eqn:flatellip}) above. Therefore, regarding the form of the metric tensor, we note three most important differences between the Kerr solution given with the form of Eq.~(\ref{eqn:metricBL}) and the Schwarzschild solution given by Eq.~(\ref{eqn:schmet}):
\begin{enumerate}
\item the reduction of the Kerr metric~(\ref{eqn:metricBL}) to flat ellipsoidal space-time~(\ref{eqn:flatellip}) by putting $M=0$ in it, results in the change of all metric tensor components (except $g_{22}$), of which one vanishes ($g_{03}$) --- for the Schwarzschild solution~(\ref{eqn:schmet}) the corresponding change occurs only in the time-time and radial-radial component,
\item the Kerr solution in the form~(\ref{eqn:metricBL}) is not orthogonal, but it does have the same symmetry as the flat ellipsoidal space-time metric~(\ref{eqn:flatellip}) which is orthogonal –-- in the case of Schwarzschild solution, the metric~(\ref{eqn:schmet}) has kept its orthogonality feature of the flat space-time counterpart~(\ref{eqn:flatsph}),
\item the product of time-time and radial-radial component of the Kerr solution in the form~(\ref{eqn:metricBL}) does not equal $-1$ --- in the case of Schwarzschild solution~(\ref{eqn:schmet}), this product does equal $-1$, just as is the case in the Minkowski metric in spherical coordinates~(\ref{eqn:flatsph}). The same difference is present between the flat metric in ellipsoidal~(\ref{eqn:flatellip}) and spherical coordinates~(\ref{eqn:flatsph}).
\end{enumerate}
Because of these differences, it seems highly unlikely that it is possible to use the same reasoning as Schwarzschild did and to ask for the ``ellipsoidally symmetric'' solution to the Einstein equations based on the metric~(\ref{eqn:flatellip}). We shall see shortly, that one could  be convinced otherwise.
These three differences vanish immediately when the Kerr metric in the form of Eq.~(\ref{eqn:metricBL}) is rewritten in the orthogonal form just by rearranging the terms to get:
\begin{equation}
\label{eqn:niceKerr}
ds^{2}=\frac{\Delta}{\rho^{2}}\left( dt-a \sin^{2}\theta d\phi \right)^{2}-\frac{\rho^{2}}{\Delta}dr^{2}-\rho^{2}d\theta^{2}-\frac{(r^{2}+a^{2})^{2}\sin^{2}\theta}{\rho^{2}}\left( d\phi-\frac{a}{r^{2}+a^{2}}dt \right)^{2}
\end{equation}
which is almost the same form as the one that can be found in Ref. (2) and Ref. (4) and also in the textbook by O\textquotesingle Neil\cite{ONeil95}, which he refers to as ``Boyer-Lindquist in orthonormal frame'', but with slightly regrouped last term:
\begin{equation}
\label{eqn:uglyKerr}
ds^{2}=\frac{\Delta}{\rho^{2}}\left( dt-a \sin^{2}\theta d\phi \right)^{2}-\frac{\rho^{2}}{\Delta}dr^{2}-\rho^{2}d\theta^{2}-\frac{\sin^{2}\theta}{\rho^{2}}\left( (r^{2}+a^{2})d\phi-a dt \right)^{2}
\end{equation}
This form can also be found in the heuristic derivation of Kerr metric by Enderlein\cite{Ender97} as a resulting metric form. We stress the importance of the Kerr solution given in the form~(\ref{eqn:niceKerr}) and we will be referring to that form in this paper.

The idea of derivation presented in the following sections is to ask for an ellipsoidally symmetric vacuum solution to the Einstein equations which doesn't possess the above stated differences with Schwarzschild solution. In other words, one searches for the metric that ``looks like'' the Schwarzschild metric~(\ref{eqn:schmet}) but for a rotating body. It turns out that one can, in principle, obtain such a solution just by proposing an orthogonal metric ansatz possessing the symmetry of ellipsoid of revolution and the same reciprocal relation between the time-time and radial-radial component found in Schwarzschild solution~(\ref{eqn:schmet}). The derivation proceeds with the use of orthonormal tetrad basis and Cartan calculus, which can be found for example in the book \emph{Mathematical Theory Of Black Holes} by Chandrasekhar\cite{Chandra83}. The result should be the Kerr metric in the form~(\ref{eqn:niceKerr}), but in this paper we only state the obtained second-order partial differential equations for the Ricci tensor in orthonormal tetrad frame and leave their final solution for a future study.

\section{The Ansatz for the metric of the Rotating Body}
A student can start the search for the metric of a rotating body by asking ``What happens to the shape of gravitational potential surfaces of a spherically symmetric non-rotating body of mass $M$ if it starts rotating?''. Due to the centrifugal force, which is strongest along the directions of $\theta=90$ deg (if the rotational axis is aligned with $z$---axis), the spherically symmetric body would deform to a rotational ellipsoid. Then the gravitational field potential in the vicinity of this body would possess an ellipsoidal symmetry.
Taking the argument to the General Relativity language, this means that the metric of curved space-time around such a body would possess ellipsoidal symmetry. Therefore, a student could ask for the solution to the Einstein equations for vacuum which possesses the ellipsoidal symmetry described by the term:
\begin{equation}
-\rho^{2}d\theta^{2}-(r^{2}+a^{2})\sin^{2}\theta d\phi^{2}
\end{equation}
from the Eq.~(\ref{eqn:flatellip}) unchanged and proceed in analogy to Eq.~(\ref{eqn:flatsphans}) with similar arguments as Schwarzschild used to find spherically symmetric solution. However, this would not be correct, because one knows that the metric of the rotating space-time would have explicitly a cross term $dtd\phi$, whereas the Minkowski metric in ellipsoidal coordinates~(\ref{eqn:flatellip}) does not have such a term, even though both metrics have the parameter $a$ which one connects to the deformation of a sphere. It seems that axial symmetry of Eq.~(\ref{eqn:flatellip}) is not manifest and exactly this fact forbids one to follow Schwarzschild's logic. Furthermore, this means that $t$ and $\phi$ are orthogonal to each other in Eq.~(\ref{eqn:flatellip}), but not in Kerr metric~(\ref{eqn:metricBL}), even though they do possess the same symmetry. Therefore, we need to get around this problem somehow, because we would like to have an unchanged manifest ellipsoidal symmetry when considering metric~(\ref{eqn:flatellip}) in the presence of mass, just as in the case with the spherically symmetric part in the Schwarzschild solution. The clue for this lies in the fact that this cross term $dtd\phi$ is hidden in the squared brackets in the orthogonal metric~(\ref{eqn:niceKerr}). Keeping this in mind, we can give Minkowski flat ellipsoidal metric~(\ref{eqn:flatellip}) the possibility of possessing the cross term $dtd\phi$ without changing its symmetry, just by hiding it by some coordinate transformation of $t$ and $\phi$ coordinate. This will ``set the stage'' for the requested ansatz for the metric of a rotating body.

One can simply guess the needed form of the new ellipsoidal flat metric by putting $M=0$ in Eq.~(\ref{eqn:uglyKerr}), but because the students ``do not yet know the solution'', this can formally be achieved by a certain coordinate transformation in the metric~(\ref{eqn:flatellip}):
\begin{eqnarray}
\label{eqn:transell}
dt\rightarrow dT & = & h dt-f d\phi \nonumber \\
\\
d\phi \rightarrow d\Phi & = & k d\phi-g dt \nonumber
\end{eqnarray}
where $h,f,k,g$ are all functions of coordinates to be determined.
Let us now write down the Minkowski metric in the ``new'' ellipsoidal coordinates, with the new metric tensor $G'_{\mu\nu}$:
\begin{equation}
ds^{2}=G'_{00}(h dt-f d\phi)^{2}+G'_{11}dr^{2}+G'_{22} d\theta^{2}+G'_{33}(k d\phi-g dt)^{2}
\end{equation}
where $G'_{11}=-\rho^{2}/(r^{2}+a^{2})$ and $G'_{22}=-\rho^{2}$ remained unchanged. Squaring out the brackets and grouping the terms to get Eq.~(\ref{eqn:flatellip}), one gets:
\begin{eqnarray}
\label{eqn:propellip}
ds^{2} & = & (G'_{00}h^{2}+G'_{33}g^{2})dt^{2}-2(G'_{00}h f+G'_{33}k g)dt d\phi \nonumber \\
& & +G'_{11}dr^{2}+G'_{22} d\theta^{2}+(G'_{00}f^{2}+G'_{33}k^{2})d\phi^{2}
\end{eqnarray}
Comparing Eq.~(\ref{eqn:propellip}) with Eq.~(\ref{eqn:flatellip}), the following system of equations should hold:
\begin{eqnarray}
\label{eqn:system}
G'_{00}h^{2} & + G'_{33}g^{2} & =1 \nonumber \\
G'_{00}h f & + G'_{33}k g & =0 \\
G'_{00}f^{2} & + G'_{33}k^{2} & =-(r^{2}+a^{2})\sin^{2}\theta \nonumber
\end{eqnarray}
This system of equations has six unknowns and only three equations, so we have three free choices. Two of them can be obtained by choosing the specific transformation~(\ref{eqn:transell}) to be simpler:
\begin{eqnarray}
\label{eqn:transells}
dt\rightarrow dT & = & dt-f d\phi \nonumber \\
\\
d\phi \rightarrow d\Phi & = & d\phi-g dt \nonumber
\end{eqnarray}
or, in other words, we choose $h=k=1$, and the Jacobian of this transformation is
\begin{equation}
\label{eqn:jacob}
J=1-f g
\end{equation}
This means that we want our new coordinates to be just a ``correction'' of the old ones---we are actually letting the change of $t$ coordinate influence the $\phi$ coordinate and vice versa, but with demand the orthogonality is preserved in the new metric. We are then left with one more degree of freedom which is the key one for the derivation of the Kerr metric in this paper. It seems that of the four unknowns $f,g,G'_{00},G'_{33}$, we can choose any to constrain, but we do not know how. However, we choose to constrain $G'_{00}$ from pure aesthetic reasons---because we want to erase the difference \#3 from page 4 between the Minkowski metric in ellipsoidal and spherical coordinates. This choice is based on the assumption that the metric~(\ref{eqn:flatellip}) can be rewritten in the form where the product of the new time-time and radial-radial component equals $-1$:
\begin{equation}
\label{eqn:reciprocal}
G'_{00}G'_{11}=-1\Rightarrow G'_{00}=-\frac{1}{G'_{11}}=-\frac{r^{2}+a^{2}}{\rho^{2}}
\end{equation}
Therefore, by doing this, one actually demands the analogous relation that holds for the Minkowski metric in spherical coordinates and for the Schwarzschild metric.
After the constraints, the system of equations~(\ref{eqn:system}) becomes:
\begin{eqnarray}
\label{eqn:system1}
-\frac{r^{2}+a^{2}}{\rho^{2}} & + G'_{33}g^{2} & =1 \nonumber \\
-\frac{r^{2}+a^{2}}{\rho^{2}} f & + G'_{33}k g & =0 \\
-\frac{r^{2}+a^{2}}{\rho^{2}}f^{2} & + G'_{33} & =-(r^{2}+a^{2})\sin^{2}\theta \nonumber
\end{eqnarray}
with exactly three unknowns, and the solutions are:
\begin{eqnarray}
\label{eqn:sols}
f & = & \pm a\sin^{2}\theta \nonumber \\
g & = & \pm \frac{a}{r^{2}+a^{2}} \\
G'_{33} & = & -\frac{(r^{2}+a^{2})^{2}\sin^{2}\theta}{\rho^{2}} \nonumber
\end{eqnarray}
where $\pm$ in $f$ and $g$ are correlated. Finally, the Minkowski metric in the ``new'' ellipsoidal coordinates reads:
\begin{eqnarray}
\label{eqn:niceell}
ds^{2} & = & \frac{r^{2}+a^{2}}{\rho^{2}}dT^{2}-\frac{\rho^{2}}{r^{2}+a^{2}}dr^{2}-\rho^{2}d\theta^{2}-\frac{(r^{2}+a^{2})^{2}\sin^{2}\theta}{\rho^{2}}d\Phi^{2} \nonumber \\
& = & \frac{r^{2}+a^{2}}{\rho^{2}}\left( dt-a \sin^{2}\theta d\phi \right)^{2}-\frac{\rho^{2}}{r^{2}+a^{2}}dr^{2}- \nonumber \\
& & \rho^{2}d\theta^{2}-\frac{(r^{2}+a^{2})^{2}\sin^{2}\theta}{\rho^{2}}\left( d\phi-\frac{a}{r^{2}+a^{2}} dt \right)^{2}
\end{eqnarray}
The Eq.~(\ref{eqn:niceell}) possesses a nice manifest connection between parameter $a$ and new coordinates $(T,\Phi)$, which can be seen when one puts $a=0$ in Eq.~(\ref{eqn:niceell}) above and spherically symmetric flat metric is restored. One can now interpret the stated connection as a proper way of introducing rotation (or manifest axial symmetry) into flat space-time metric---by introducing the change of both $t$ and $\phi$ coordinate, not just $\phi$.
At this point one can observe the power of this coordinate transformation. Firstly, one demands that new coordinates $T$ and $\Phi$ are orthogonal to each other. Secondly, one demands that dimension of $T$ coordinate is the same as dimension of $t$ coordinate (time) and the same demand is put on $\Phi$ coordinate (dimension of an angle), which provides a physical argument for introducing the coordinate transformation in the form of~(\ref{eqn:transells}). Thirdly, one demands the new metric to be orthogonal and employs a rather aesthetic and ``Schwarzschild metric look-like'' argument for the constrain on $G'_{00}$, by demanding reciprocity relation~(\ref{eqn:reciprocal}) to hold for the new metric tensor elements. This ``aesthetic'' argument actually has its roots in an interesting feature of vacuum solutions, which says that for the 4-velocity of radial null curve $k^{\mu}$, the energy---momentum tensor satisfies $ T_{\mu\nu}k^{\mu}k^{\nu}=0 $, which is explained in a paper by Jacobson\cite{Jacob07}. One could equivalently have a requirement that radially in-falling photon does not feel acceleration and then the reciprocity relation~(\ref{eqn:reciprocal}) also follows, as pointed by Dadhich\cite{Dadhich12} and Jacobson\cite{Jacob07}. The latter requirement is physically more intuitive for a student to observe. As a result, one comes at the metric of ellipsoidal space-time~(\ref{eqn:niceell}) which, when compared to Eq.~(\ref{eqn:niceKerr}), differs only in the ``new'' time-time and radial-radial metric tensor components, just as is the case with spherically symmetric analogues. Therefore, one can say that the fact that Kerr metric can be written in the form of~(\ref{eqn:uglyKerr}) or~(\ref{eqn:niceKerr}) is a consequence of the existence of a coordinate transformation~(\ref{eqn:transells}) which naturally introduces rotation in flat space-time by ``generating'' the cross term $dt d\phi$.
This cross term vanishes in flat space, when the brackets in Eq.~(\ref{eqn:niceell}) are squared. But what if we wanted to find the solution to the Einstein equations for vacuum which has the same manifest symmetry as Eq.~(\ref{eqn:niceell})? We actually want to find a metric of the space-time curved by the same mass $M$ as in Schwarzschild metric, but which is rotating. This means that flat metric~(\ref{eqn:niceell}) would change in such a way to produce the non-vanishing cross term $dt d\phi$. Phenomenologically speaking only, this can only be achieved with the change of either $G'_{00}$ or $G'_{33}$ (or both) to some new functions $G_{TT}$ and $G_{\Phi\Phi}$ so that the corresponding cross terms do not cancel. In this way, by searching for the new functions $G_{TT}$ or $G_{\Phi\Phi}$ (or both) as solutions to the Einstein equations one actually searches for a mathematically valid way of introducing rotation into metric~(\ref{eqn:flatellip}) and obtaining the metric of a rotating body.
Now, we could proceed in two ways. One is to use an ansatz:
\begin{eqnarray}
\label{eqn:kerransatz}
ds^{2} & = & e^{2\nu(r,\theta)}(dt-a\sin^{2}\theta d\phi)^{2}+e^{2\sigma(r,\theta)}dr^{2}- \nonumber \\
& & (r^{2}+a^{2}\cos^{2}\theta)d\theta^{2}-\frac{(r^{2}+a^{2})^{2}\sin^{2}\theta}{\rho^{2}}\left( d\phi-\frac{a}{r^{2}+a^{2}} dt \right)^{2}
\end{eqnarray}
which would be analogous to what Schwarzschild did to get spherically symmetric solution. This ansatz is set with the requirement that the presence of a rotating mass preserves ellipsoidal symmetry described by the following term:
\begin{equation}
-(r^{2}+a^{2}\cos^{2}\theta)d\theta^{2}-\frac{(r^{2}+a^{2})^{2}\sin^{2}\theta}{\rho^{2}}\left( d\phi-\frac{a}{r^{2}+a^{2}} dt \right)^{2}
\end{equation}
It would be interesting to check whether it is possible to derive the Kerr metric from the ansatz~(\ref{eqn:kerransatz}), but we leave that for some future paper.
Instead, we will treat all of the ``new'' metric tensor components as general functions and simply search for the solution starting from an ansatz:
\begin{equation}
\label{eqn:kerransatz1}
ds^{2}=e^{2\nu}(dt-a\sin^{2}\theta d\phi)^{2}-e^{2\mu}\left( d\phi-\frac{a}{r^{2}+a^{2}} dt \right)^{2}-e^{-2\nu}dr^{2}-e^{2\lambda}d\theta^{2}
\end{equation}
where $\nu,\lambda,\mu$ are all functions of $r$ and $\theta$. We have used the following demands to form an ansatz~(\ref{eqn:kerransatz1}) for the rotating body:
\begin{itemize}
\item the reciprocity relation~(\ref{eqn:reciprocal}), because we want to search for the solution that does not have the difference $\#3$ on page 4 with Schwarzschild solution
\item we have used the coordinate transformation~(\ref{eqn:transells}) because we want to search for the solution that---as in the case of Schwarzschild solution---features the orthogonality of the flat space-time counterpart~(\ref{eqn:niceell}), where the new coordinates and metric tensor components have the same dimension as the old ones and thereby erasing the differences $\#1$ and $\#2$ on page 3 with Schwarzschild solution.
\end{itemize}
In this way, all of the differences between the yet to be found Kerr metric form and Schwarzschild metric~(\ref{eqn:schmet}) stated in the introduction would be removed. Therefore, we force the ansatz to obey the same characteristics as the Schwarzschild metric, by using one degree of gauge freedom as the reciprocity relation~(\ref{eqn:reciprocal}) and then try to solve Einstein equations for vacuum to find the solution. Moreover, we are left with only three unknown functions $e^{2\nu}$, $e^{2\lambda}$ and $e^{2\mu}$  to find.

\section{Employing Cartan Calculus to Find the Riemann and Ricci Tensor}
We use Cartan calculus of differential forms (presented for example in Ref. (2), p. 10---40) to find the Ricci tensor for the ansatz~(\ref{eqn:kerransatz1}). First, we will introduce the notation:
\begin{equation}
dx^{0}=dt \hspace{1cm} dx^{1}=d\phi \hspace{1cm} dx^{2}=dr \hspace{1cm} dx^{3}=d\theta
\end{equation}
and also:
\begin{equation}
\label{eqn:fg}
f\equiv a\sin^{2}\theta \hspace{1cm} g\equiv \frac{a}{r^{2}+a^{2}}
\end{equation}
Based on this and the ansatz~(\ref{eqn:kerransatz1}), we have the transformation to the orthonormal tetrad basis:
\begin{eqnarray}
\label{eqn:basisforms}
\omega^{0} & = & e^{\nu}(dx^{0}-f dx^{1}) \nonumber \\
\omega^{1} & = & e^{\mu}(dx^{1}-g dx^{0}) \nonumber \\
\omega^{2} & = & e^{-\nu}dx^{2} \nonumber \\
\omega^{3} & = & e^{\lambda}dx^{3}
\end{eqnarray}
with the metric $\eta_{ab}=diag(1,-1,-1,-1)$ for rising/lowering indices. We are going to need the inverse transformation also:
\begin{eqnarray}
\label{eqn:basiscoord}
dx^{0} & = & X(e^{-\nu}\omega^{0}+f e^{-\mu}\omega^{1}) \nonumber \\
dx^{1} & = & X(e^{-\mu}\omega^{1}+g e^{-\nu}\omega^{0}) \nonumber \\
dx^{2} & = & e^{\nu}\omega^{2} \nonumber \\
dx^{3} & = & e^{-\lambda}\omega^{3} 
\end{eqnarray}
where $X\equiv J^{-1}$ is the inverse of Jacobian~(\ref{eqn:jacob}). In order to find the Ricci tensor, one first has to find the Riemann tensor and this is accomplished by the use of the first Cartan equation (Ref. 2, eq. 137, p. 22):
\begin{equation}
\label{eqn:frstcart}
d\omega^{a}=-{\omega^{a}}_{b}\wedge\omega^{b}
\end{equation}
where ${\omega^{a}}_{b}$ are connection 1---forms and the second Cartan equation (Ref. 2, eq. 148, p. 23):
\begin{equation}
\label{eqn:seccart}
\frac{1}{2}{R^{a}}_{bcd}\omega^{c}\wedge\omega^{d}=d{\omega^{a}}_{b}+{\omega^{a}}_{c}\wedge{\omega^{c}}_{b}\equiv{\Omega^{a}}_{b}
\end{equation}
where ${R^{a}}_{bcd}$ is Riemann tensor in orthonormal tetrad basis.
First we make an exterior differentiation of 1-forms~(\ref{eqn:basisforms}), with the help of~(\ref{eqn:basiscoord}):
\begin{eqnarray}
d\omega^{0} & = & d\left[e^{\nu}(dx^{0}-f dx^{1})\right] \nonumber \\
& = & -e^{\nu}\left( \nu_{,2}-f g \nu_{,2}-g f_{,2}\right) X \omega^{0}\wedge\omega^{2}-e^{-\lambda}\left( \nu_{,3}-f g \nu_{,3}-g f_{,3}\right) X \omega^{0}\wedge\omega^{3} \nonumber \\
& & + f_{,2}e^{2\nu-\mu} X \omega^{1}\wedge\omega^{2}+f_{,3}e^{\nu-\lambda-\mu} X \omega^{1}\wedge\omega^{3} \\
d\omega^{1} & = & d\left[e^{\mu}(dx^{1}-g dx^{0})\right] \nonumber\\
& = & -e^{\nu}\left( \mu_{,2}-f g \mu_{,2}-f g_{,2}\right) X \omega^{1}\wedge\omega^{2}-e^{-\lambda}\left( \mu_{,3}-f g \mu_{,3}-f g_{,3}\right) X \omega^{1}\wedge\omega^{3}\nonumber \\
& & + g_{,2}e^{\mu} X \omega^{0}\wedge\omega^{2}+g_{,3}e^{\mu-\lambda-\nu} X \omega^{0}\wedge\omega^{3} \\
d\omega^{2} & = & d\left( e^{-\nu}dx^{2}\right)=\nu_{,3}e^{-\lambda}\omega^{2}\wedge\omega^{3} \\
d\omega^{3} & = & d\left( e^{\lambda}dx^{3}\right) =-\lambda_{,3}e^{\nu}\omega^{3}\wedge\omega^{1}
\end{eqnarray}
where we use the notation $F_{,i}$ for the partial derivative of any function $F=\nu,\mu,f,g$ w.r.t. $i$---th coordinate. Then we use first Cartan equation~(\ref{eqn:frstcart}) to identify the connection 1-forms ${\omega^{a}}_{b}$:
\begin{eqnarray}
{\omega^{0}}_{2} & = & e^{\nu}\left( \nu_{,2}-f g \nu_{,2}-g f_{,2}\right) X \omega^{0}-f_{,2}e^{2\nu-\mu} X \omega^{1} \\
{\omega^{0}}_{3} & = & e^{-\lambda}\left( \nu_{,3}-f g \nu_{,3}-g f_{,3}\right) X \omega^{0}-f_{,3}e^{\nu-\lambda-\mu} X \omega^{1} \\
{\omega^{1}}_{2} & = & e^{\nu}\left( \mu_{,2}-f g \mu_{,2}-f g_{,2}\right) X \omega^{1}-g_{,2}e^{\mu} X \omega^{0} \\
{\omega^{1}}_{3} & = & e^{-\lambda}\left( \mu_{,3}-f g \mu_{,3}-f g_{,3}\right) X \omega^{1}-g_{,3}e^{\mu-\lambda-\nu} X \omega^{0} \\
{\omega^{2}}_{3} & = & -\nu_{,3}e^{-\lambda}\omega^{2}-\lambda_{,2}e^{\nu}\omega^{3}
\end{eqnarray}
Differentiation of the above connection one-forms leads us to:
\begin{eqnarray}
\label{eqn:difonebeg}
d{\omega^{0}}_{2} & = & d\left[ e^{\nu}\left( \nu_{,2}-f g \nu_{,2}-g f_{,2}\right) X \omega^{0}-f_{,2}e^{2\nu-\mu} X \omega^{1} \right] \nonumber \\
& = & \left\lbrace 
\left[ e^{2\nu} \left( f\nu_{,2}+f_{,2} \right) \right]_{,2}g-\left(e^{2\nu}\nu_{,2}\right)_{,2}
\right\rbrace X\omega^{0}\wedge\omega^{2} \nonumber \\
& + & \left\lbrace e^{-(\lambda+\nu)}\left[ e^{2\nu}\left(f \nu_{,2}+f_{,2}\right)\right]_{,3}g-e^{-(\lambda+\nu)}\left(e^{2\nu}\nu_{,2}\right)_{,3}\right\rbrace X\omega^{0}\wedge\omega^{3} \nonumber \\
& + & \left\lbrace e^{\nu-\mu}\left[ e^{2\nu}\left(f \nu_{,2}+f_{,2}\right)\right]_{,2}-e^{\nu-\mu}\left(e^{2\nu}\nu_{,2}\right)_{,2}f\right\rbrace X\omega^{1}\wedge\omega^{2} \nonumber \\
& + & \left\lbrace e^{-(\lambda+\mu)}\left[ e^{2\nu}\left(f \nu_{,2}+f_{,2}\right)\right]_{,3}-e^{-(\lambda+\mu)}\left(e^{2\nu}\nu_{,2}\right)_{,3}f\right\rbrace X\omega^{1}\wedge\omega^{3}
\end{eqnarray}
\begin{eqnarray}
d{\omega^{0}}_{3} & = & d\left[ e^{-\lambda}\left( \nu_{,3}-f g \nu_{,3}-g f_{,3}\right) X \omega^{0}-f_{,3}e^{\nu-\lambda-\mu} X \omega^{1} \right] \nonumber \\
& = & \left\lbrace 
\left[ e^{\nu-\lambda} \left( f\nu_{,3}+f_{,3} \right) \right]_{,2}g-\left(e^{\nu-\lambda}\nu_{,3}\right)_{,2}
\right\rbrace X\omega^{0}\wedge\omega^{2} \nonumber \\
& + & \left\lbrace e^{-(\lambda+\nu)}\left[ e^{\nu-\lambda}\left(f \nu_{,3}+f_{,3}\right)\right]_{,3}g-e^{-(\lambda+\nu)}\left(e^{\nu-\lambda}\nu_{,3}\right)_{,3}\right\rbrace X\omega^{0}\wedge\omega^{3} \nonumber \\
& + & \left\lbrace e^{\nu-\mu}\left[ e^{\nu-\lambda}\left(f \nu_{,3}+f_{,3}\right)\right]_{,2}-e^{\nu-\mu}\left(e^{\nu-\lambda}\nu_{,3}\right)_{,2}f\right\rbrace X\omega^{1}\wedge\omega^{2} \nonumber \\
& + & \left\lbrace e^{-(\lambda+\mu)}\left[ e^{\nu-\lambda}\left(f \nu_{,3}+f_{,3}\right)\right]_{,3}-e^{-(\lambda+\mu)}\left(e^{\nu-\lambda}\nu_{,3}\right)_{,3}f\right\rbrace X\omega^{1}\wedge\omega^{3}
\end{eqnarray}
\begin{eqnarray}
d{\omega^{1}}_{2} & = & d\left[ e^{\nu}\left( \mu_{,2}-f g \mu_{,2}-f g_{,2}\right) X \omega^{1}-g_{,2}e^{\mu} X \omega^{0} \right] \nonumber \\
& = & \left\lbrace 
\left[ e^{\nu+\mu} \left( g\mu_{,2}+g_{,2} \right) \right]_{,2}-\left(e^{\nu+\mu}\mu_{,2}\right)_{,2}g
\right\rbrace X\omega^{0}\wedge\omega^{2} \nonumber \\
& + & \left\lbrace e^{-(\lambda+\nu)}\left[ e^{\nu+\mu}\left(g \mu_{,2}+g_{,2}\right)\right]_{,3}-e^{-(\lambda+\nu)}\left(e^{\nu+\mu}\mu_{,2}\right)_{,3}g\right\rbrace X\omega^{0}\wedge\omega^{3} \nonumber \\
& + & \left\lbrace e^{\nu-\mu}\left[ e^{\nu+\mu}\left(g \mu_{,2}+g_{,2}\right)\right]_{,2}f-e^{\nu-\mu}\left(e^{\nu+\mu}\mu_{,2}\right)_{,2}\right\rbrace X\omega^{1}\wedge\omega^{2} \nonumber \\
& + & \left\lbrace e^{-(\lambda+\mu)}\left[ e^{\nu+\mu}\left(g \mu_{,2}+g_{,2}\right)\right]_{,3}f-e^{-(\lambda+\mu)}\left(e^{\nu+\mu}\mu_{,2}\right)_{,3}\right\rbrace X\omega^{1}\wedge\omega^{3}
\end{eqnarray}
\begin{eqnarray}
d{\omega^{1}}_{3} & = & d\left[ e^{-\lambda}\left( \mu_{,3}-f g \mu_{,3}-f g_{,3}\right) X \omega^{1}-g_{,3}e^{\mu-\lambda-nu} X \omega^{0} \right] \nonumber \\
& = & \left\lbrace 
\left[ e^{\mu-\lambda} \left( g\mu_{,3}+g_{,3} \right) \right]_{,2}-\left(e^{\mu-\lambda}\mu_{,3}\right)_{,2}g
\right\rbrace X\omega^{0}\wedge\omega^{2} \nonumber \\
& + & \left\lbrace e^{-(\lambda+\nu)}\left[ e^{\mu-\lambda}\left(g\mu_{,3}+g_{,3} \right)\right]_{,3}-e^{-(\lambda+\nu)}\left(e^{\mu-\lambda}\mu_{,3}\right)_{,3}g\right\rbrace X\omega^{0}\wedge\omega^{3} \nonumber \\
& + & \left\lbrace e^{\nu-\mu}\left[ e^{\mu-\lambda}\left(g\mu_{,3}+g_{,3}\right)\right]_{,2}f-e^{\nu-\mu}\left(e^{\mu-\lambda}\mu_{,3}\right)_{,2}\right\rbrace X\omega^{1}\wedge\omega^{2} \nonumber \\
& + & \left\lbrace e^{-(\lambda+\mu)}\left[ e^{\mu-\lambda}\left(g\mu_{,3}+g_{,3}\right)\right]_{,3}f-e^{-(\lambda+\mu)}\left(e^{\mu-\lambda}\mu_{,3}\right)_{,3}\right\rbrace X\omega^{1}\wedge\omega^{3}
\end{eqnarray}
\begin{eqnarray}
\label{eqn:difoneend}
{d\omega^{2}}_{3} & = & -d\left( \nu_{,3}e^{-\lambda}\omega^{2}+\lambda_{,2}e^{\nu}\omega^{3}\right)=\left\lbrace \left( e^{\nu+\lambda}\lambda_{,2}\right)_{,2}-\left( e^{-(\nu+\lambda)}\nu_{,3}\right)_{,3}\right\rbrace \omega^{3}\wedge\omega^{2}
\end{eqnarray}
Using Eqns.~(\ref{eqn:difonebeg})---~(\ref{eqn:difoneend}) in the right hand side of the second Cartan equation~(\ref{eqn:seccart}), one is able to read the components of the Riemann tensor ${R^{a}}_{bcd}$:
\begin{eqnarray}
{R^{0}}_{101} & = & e^{2\nu}\left[ X\left( g f_{,2}\mu_{,2}+f g_{,2}\nu_{,2}-g_{,2}f_{,2} \right)-\mu_{,2}\nu_{,2}\right] \nonumber \\
& + & e^{-2\lambda}\left[ X\left( g f_{,3}\mu_{,3}+f g_{,3}\nu_{,3}-g_{,3}f_{,3} \right)-\mu_{,3}\nu_{,3}\right]
\end{eqnarray}
\begin{eqnarray}
{R^{0}}_{202} & = & \left\lbrace \left[ e^{2\nu}\left(f\nu_{,2}+f_{,2} \right)\right]_{,2}g-\left( e^{2\nu}\nu_{,2}\right)_{,2}+\nu_{,3}e^{-2\lambda}\left( \nu_{,3}-f g \nu_{,3}-g f_{,3}\right)\right\rbrace X\\
{R^{0}}_{212} & = & e^{\nu-\mu}\left\lbrace \left[ e^{2\nu}\left(f\nu_{,2}+f_{,2} \right)\right]_{,2}-\left(e^{2\nu}\nu_{,2}\right)_{,2}f-f_{,3}\nu_{,3}e^{-2\lambda} \right\rbrace X\\
{R^{0}}_{203} & = & e^{-\lambda+\nu}\left\lbrace \left[ e^{2\nu}\left(f\nu_{,2}+f_{,2} \right)\right]_{,3}g-\left( e^{2\nu}\nu_{,2}\right)_{,3}+\lambda_{,2}e^{2\nu}\left( \nu_{,3}-f g \nu_{,3}-g f_{,3}\right)\right\rbrace X\\
{R^{0}}_{213} & = & e^{-(\lambda+\mu)}\left\lbrace \left[ e^{2\nu}\left(f\nu_{,2}+f_{,2} \right)\right]_{,3}-\left(e^{2\nu}\nu_{,2}\right)_{,3}f-f_{,3}\lambda_{,3}e^{2\nu} \right\rbrace X
\end{eqnarray}
\begin{eqnarray}
{R^{0}}_{302} & = & \left\lbrace \left[ e^{\nu-\lambda}\left(f\nu_{,3}+f_{,3} \right)\right]_{,2}g\right.\nonumber\\
& & \left.-\left( e^{\nu-\lambda}\nu_{,3}\right)_{,2}-\nu_{,3}e^{\nu-\lambda}\left( \nu_{,2}-f g \nu_{,2}-g f_{,2}\right)\right\rbrace X
\label{eqn:R0312}\\
{R^{0}}_{312} & = & e^{\nu-\mu}\left\lbrace \left[ e^{\nu-\lambda}\left(f\nu_{,3}+f_{,3} \right)\right]_{,2}-\left(e^{\nu-\lambda}\nu_{,3}\right)_{,2}f+f_{,2}\nu_{,3}e^{\nu-\lambda} \right\rbrace X\\
{R^{0}}_{303} & = & e^{-(\lambda+\nu)}\left\lbrace \left[ e^{\nu-\lambda}\left(f\nu_{,3}+f_{,3} \right)\right]_{,3}g-\left( e^{\nu-\lambda}\nu_{,3}\right)_{,3}\right.\nonumber \\
& & \left. -\lambda_{,2}e^{2\nu}\left( \nu_{,2}-f g \nu_{,2}-g f_{,2}\right)\right\rbrace X\\
{R^{0}}_{313} & = & e^{-(\lambda+\mu)}\left\lbrace \left[ e^{\nu-\lambda}\left(f\nu_{,3}+f_{,3} \right)\right]_{,3}-\left(e^{\nu-\lambda}\nu_{,3}\right)_{,3}f+f_{,2}\lambda_{,2}e^{2\nu} \right\rbrace X
\end{eqnarray}
\begin{eqnarray}
{R^{1}}_{202} & = & \left\lbrace \left[ e^{\nu+\mu}\left(g\mu_{,2}+g_{,2} \right)\right]_{,2}-\left(e^{\nu+\mu}\mu_{,2}\right)_{,2}g-g_{,3}\nu_{,3}e^{\mu-\lambda-\nu} \right\rbrace X\\
{R^{1}}_{212} & = & e^{\nu-\mu}\left\lbrace \left[ e^{\nu+\mu}\left(g\mu_{,2}+g_{,2} \right)\right]_{,2}f-\left( e^{\nu+\mu}\mu_{,2}\right)_{,2}\right.\nonumber\\
& & \left. +\nu_{,3}e^{\mu-\nu-2\lambda}\left( \mu_{,3}-f g \mu_{,3}-f g_{,3}\right)\right\rbrace X\\
{R^{1}}_{203} & = & e^{-(\lambda+\nu)}\left\lbrace \left[ e^{\nu+\mu}\left(g\mu_{,2}+g_{,2} \right)\right]_{,3}-\left(e^{\nu+\mu}\mu_{,2}\right)_{,3}g-g_{,3}\lambda_{,2}e^{\nu+\mu} \right\rbrace X\\
{R^{1}}_{213} & = & e^{-(\nu+\mu)}\left\lbrace \left[ e^{\nu+\mu}\left(g\mu_{,2}+g_{,2} \right)\right]_{,3}f-\left( e^{\nu+\mu}\mu_{,2}\right)_{,3}\right.\nonumber\\
& & \left. +\lambda_{,2}e^{2\nu}\left( \mu_{,3}-f g \mu_{,3}-f g_{,3}\right)\right\rbrace X
\end{eqnarray}
\begin{eqnarray}
{R^{1}}_{302} & = & \left\lbrace \left[ e^{\mu-\lambda}\left(g\mu_{,3}+g_{,3} \right)\right]_{,2}-\left(e^{\mu-\lambda}\mu_{,3}\right)_{,2}g+g_{,2}\nu_{,3}e^{\nu-\lambda} \right\rbrace X\\
{R^{1}}_{312} & = & e^{\nu-\mu}\left\lbrace \left[ e^{\mu-\lambda}\left(g\mu_{,3}+g_{,3} \right)\right]_{,2}f-\left( e^{\mu-\lambda}\mu_{,3}\right)_{,2}\right.\nonumber\\
& & \left. -\nu_{,3}e^{\mu-\lambda}\left( \mu_{,2}-f g \mu_{,2}-f g_{,2}\right)\right\rbrace X\\
{R^{1}}_{303} & = & e^{-(\lambda+\nu)}\left\lbrace \left[ e^{\mu-\lambda}\left(g\mu_{,3}+g_{,3} \right)\right]_{,3}-\left(e^{\mu-\lambda}\mu_{,3}\right)_{,3}g+g_{,2}\lambda_{,2}e^{\mu+\lambda} \right\rbrace X\\
{R^{1}}_{313} & = & e^{-(\lambda+\mu)}\left\lbrace \left[ e^{\mu-\lambda}\left(g\mu_{,3}+g_{,3} \right)\right]_{,3}f-\left( e^{\mu-\lambda}\mu_{,3}\right)_{,3}\right.\nonumber\\
& & \left.-\lambda_{,2}e^{2\nu+\lambda+\mu}\left( \mu_{,2}-f g \mu_{,2}-f g_{,2}\right)\right\rbrace X
\end{eqnarray}
\begin{eqnarray}
{R^{2}}_{323} & = & -e^{\nu-\lambda}\left\lbrace \left[e^{\nu}\left(e^{\lambda} \right)_{,2} \right]_{,2}+\left[e^{-\lambda}\left(e^{\nu} \right)_{,3} \right]_{,3}\right\rbrace\\
{R^{2}}_{301} & = & e^{\mu-\lambda}\left\lbrace \left( g_{,2}\mu_{,3}-g_{,3}\mu_{,2}\right)+e^{\nu}\left(f_{,3}\nu_{,2}-f_{,2}\nu_{,3} \right) \right\rbrace X
\end{eqnarray}
We have arrived at the equations for 19 components of the Riemann tensor, but they are not all independent, because of the symmetry reasons and some identities. For example, Jacobi identity gives us the following relation:
\begin{equation}
{R^{0}}_{123}+{R^{0}}_{231}+{R^{0}}_{312}=0
\end{equation}
Now, the component ${R^{0}}_{312}$ is present in Eq.~(\ref{eqn:R0312}) and we can derive the remaining two based on symmetry and antisymmetry of the Riemann tensor:
\begin{eqnarray}
{R^{0}}_{213} & = & -{R^{0}}_{231}\\
{R^{2}}_{301} & = & -{R^{0}}_{123}
\end{eqnarray}
so the Jacobi identity becomes:
\begin{equation}
-{R^{2}}_{301}-{R^{0}}_{213}+{R^{0}}_{312}=0
\end{equation}
which could be useful relation. Also, there are three more relations:
\begin{eqnarray}
{R^{0}}_{212} & = & -{R^{1}}_{202}\\
{R^{0}}_{313} & = & -{R^{1}}_{303}\\
{R^{1}}_{213} & = & {R^{1}}_{312}
\end{eqnarray}
which further reduce the number of independent components of Riemann tensor to 13. Finally, Ricci tensor is derived by contracting the Riemann tensor (we also use some symmetry relations of the Riemann tensor):
\begin{eqnarray}
R_{00} & = & {R^{a}}_{0a0}={R^{1}}_{010}+{R^{2}}_{020}+{R^{3}}_{030}=-\left({R^{1}}_{010}+{R^{2}}_{020}+{R^{3}}_{030}\right)\\
R_{11} & = & {R^{a}}_{1a1}={R^{0}}_{101}+{R^{2}}_{121}+{R^{3}}_{131}\\
R_{22} & = & {R^{a}}_{2a2}={R^{0}}_{202}+{R^{1}}_{212}+{R^{3}}_{232}={R^{0}}_{202}+{R^{1}}_{212}+{R^{2}}_{323}\\
R_{33} & = & {R^{a}}_{3a3}={R^{0}}_{303}+{R^{1}}_{313}+{R^{2}}_{323}\\
R_{01} & = & {R^{a}}_{0a1}={R^{0}}_{212}+{R^{0}}_{313}\\
R_{23} & = & {R^{a}}_{2a3}={R^{0}}_{203}+{R^{1}}_{213}
\end{eqnarray}
Setting the Einstein equations for the vacuum:
\begin{equation}
R_{ab}-\frac{1}{2}\eta_{ab}R=0
\end{equation}
it follows that $R=0$ and  $R_{ab}=0$. Then one finally has the equations for Ricci tensor:
\begin{eqnarray}
-R_{00} & = & e^{2\nu}\left[X\left( g f_{,2}\mu_{,2}+f g_{,2}\nu_{,2}-g_{,2}f_{,2}\right)-\mu_{,2}\nu_{,2}\right]\nonumber \\
& & + e^{-2\lambda}\left[X \left( g f_{,3}\mu_{,3}+f g_{,3}\nu_{,3}-g_{,3}f_{,3}\right)-\mu_{,3}\nu_{,3}\right]\nonumber \\
& & + \left\lbrace \left[e^{2\nu} \left(f\nu_{,2}+f_{,2} \right) \right]_{,2} g-\left(e^{2\nu} \nu_{,2} \right)_,2+\nu_,3 e^{-2\lambda} \left(\nu_{,3}-f g \nu_{,3}-g f_{,3} \right) \right\rbrace \nonumber\\
& & +e^{-(\lambda+\nu)} \left\lbrace \left[e^{\nu-\lambda} \left(f\nu_{,3}+f_{,3} \right) \right]_{,3} g-\left(e^{\nu-\lambda} \nu_{,3} \right)_{,3}\right.\nonumber\\
& & \left.-\lambda_{,2} e^{2\nu}\left(\nu_{,2}-f g \nu_{,2}-g f_{,2} \right) \right\rbrace X = 0
\end{eqnarray}
\begin{eqnarray}
R_{11} & = & e^{2\nu} \left[X\left(gf_{,2} \mu_{,2}+f g_{,2} \nu_{,2}-g_{,2} f_{,2} \right)-\mu_{,2} \nu_{,2} \right]\nonumber \\
& & +e^{-2\lambda} \left[ X\left( g f_{,3} \mu_{,3}+f g_{,3} \nu_{,3}-g_{,3} f_{,3} \right)-\mu_{,3} \nu_{,3} \right]\nonumber \\
& & +e^{\nu-\mu} \left\lbrace \left[e^{\nu+\mu} \left(g\mu_{,2}+g_{,2} \right) \right]_{,2} f-\left(e^{\nu+\mu} \mu_{,2} \right)_{,2}\right.\nonumber\\
& & \left.+\nu_{,3} e^{\mu-nu-2\lambda} \left(\mu_{,3}-f g \mu_{,3}-f g_{,3} \right) \right\rbrace X \nonumber \\
& & +e^{-(\lambda+\mu)} \left\lbrace \left[e^{\mu-\lambda} \left( g\mu_{,3}+g_{,3} \right) \right]_{,3} f-\left(e^{\mu-\lambda} \mu_{,3} \right)_{,3}\right.\nonumber\\
& & \left. -\lambda_{,2} e^{2\nu+\lambda+\mu}\left(\mu_{,2}-f g \mu_{,2}-f g_{,2} \right) \right\rbrace X = 0
\end{eqnarray}
\begin{eqnarray}
R_{22} & = & \left\lbrace \left[e^{2\nu} \left(f \nu_{,2}+f_{,2} \right) \right]_{,2} g-\left(e^{2\nu} \nu_{,2} \right)_{,2}+\nu_{,3} e^{-2\lambda} \left(\nu_{,3}-f g \nu_{,3}-g f_{,3} \right) \right\rbrace X \nonumber \\
& & +e^{\nu-\mu} \left\lbrace \left[e^{\nu+\mu} \left(g \mu_{,2}+g_{,2} \right) \right]_{,2} f-\left(e^{\nu+mu} \mu_{,2} \right)_{,2}\right.\nonumber\\
& & \left.+\nu_{,3} e^{\mu-\nu-2\lambda} \left(\nu_,3-f g \mu_{,3}-f g_{,3} \right) \right\rbrace X \nonumber \\
& & -e^{\nu-\lambda} \left\lbrace \left[e^{\nu} \left(e^{\lambda} \right)_{,2} \right]_{,2}+\left[e^{-\lambda} \left(e^{\nu} \right)_{,3} \right]_{,3} \right\rbrace = 0
\end{eqnarray}
\begin{eqnarray}
R_{33} & = & e^{-(\lambda+\nu)} \left\lbrace \left[e^{\nu-\lambda} \left(f\nu_{,3}+f_{,3} \right) \right]_{,3} g-\left(e^{\nu-\lambda} \nu_{,3} \right)_{,3}-\lambda_{,2} e^{2\nu} \left(\nu_{,2}-f g \nu_{,2}-g f_{,2} \right) \right\rbrace X\nonumber \\
& & +e^{-(\lambda+\nu)} \left\lbrace \left[e^{\mu-\lambda} \left(g \mu_{,3}+g_{,3} \right) \right]_{,3} f-\left(e^{\mu-\lambda} \mu_{,3} \right)_{,3}\right.\nonumber\\
& & \left.-\lambda_{,2} e^{2\nu+\mu+\lambda} \left(\mu_{,2}-f g \mu_{,2}-f g_{,2} \right) \right\rbrace X \nonumber \\
& & -e^{\nu-\lambda} \left\lbrace \left[e^{\nu}\left(e^{\lambda} \right)_{,2} \right]_{,2}+\left[e^{-\lambda} \left(e^{\nu} \right)_{,3} \right]_{,3} \right\rbrace =0
\end{eqnarray}
\begin{eqnarray}
R_{01} & = & e^{\nu-\mu} \left\lbrace \left[e^{2\nu} \left(f\nu_{,2}+f_{,2} \right) \right]_{,2}-\left(e^{2\nu} \nu_{,2} \right)_{,2} f-f_{,3} \nu_{,3} e^{-2\lambda} \right\rbrace X \nonumber \\
& & +e^{-(\lambda+\mu)} \left\lbrace \left[e^{\nu-\lambda} \left(f\nu_{,3}+f_{,3} \right) \right]_{,3}-\left(e^{\nu-\lambda} \nu_{,3} \right)_{,3} f+f_{,2} \lambda_{,2} e^{2\nu} \right\rbrace X=0
\end{eqnarray}
\begin{eqnarray}
R_{23} & = & e^{-(\lambda+\nu)} \left\lbrace \left[e^{2\nu} \left(f\nu_{,2}+f_{,2} \right) \right]_{,3} g-\left(e^{\nu} \nu_{,2} \right)_{,3}+\lambda_{,2} e^{2\nu} \left(\nu_{,3}-f g \nu_{,3}-g f_{,3} \right) \right\rbrace X \nonumber \\
& & +e^{-(\mu+\nu)} \left\lbrace \left[e^{\mu+\nu} \left(g \mu_{,2}+g_{,2} \right) \right]_{,3} f-\left(e^{\mu+\nu} \mu_{,2} \right)_{,3}\right.\nonumber\\
& & \left.+\lambda_{,2} e^{2\nu} \left(\mu_{,3}-f g \mu_{,3}-f g_{,3} \right) \right\rbrace  X = 0
\end{eqnarray}
These equations could be solved in some future paper. One bears in mind that the functions $f$ and $g$ are known from Eq.~(\ref{eqn:fg}) and that there are only three unknown functions which have to be found: $e^{2\nu}$, $e^{2\lambda}$ and $e^{2\mu}$. Also, since we are trying to find the solution which we want to be reduced to the Schwarzschild solution when the body stops rotating, then one can use the following limits too:
\begin{eqnarray}
\lim\limits_{a\rightarrow 0}e^{2\nu} & = & \frac{r^{2}-2Mr}{r^{2}}\nonumber \\
\lim\limits_{a\rightarrow 0}e^{2\lambda} & = & r^{2} \\
\lim\limits_{a\rightarrow 0}e^{2\mu} & = & r^{2}\sin^{2}\theta \nonumber
\end{eqnarray}
One could also demand that in the limit of flat space-time, the solution reduces to Eq.~(\ref{eqn:niceell}). The Riemann tensor components then tend to zero and the unknown functions tend to metric tensor components in Eq.~(\ref{eqn:niceell}):
\begin{eqnarray}
\lim\limits_{a\rightarrow 0}e^{2\nu} & = & \frac{r^{2}+a^{2}}{r^{2}+a^{2}\cos^{2}\theta}\nonumber \\
\lim\limits_{a\rightarrow 0}e^{2\lambda} & = & r^{2}+a^{2}\cos^{2}\theta \\
\lim\limits_{a\rightarrow 0}e^{2\mu} & = & \frac{(r^{2}+a^{2})^{2}\sin^{2}\theta}{r^{2}+a^{2}\cos^{2}\theta} \nonumber
\end{eqnarray}
This could maybe be useful when trying to solve the equations of Ricci tensor.
Finally, one should be able to get the metric:
\begin{eqnarray}
ds^{2} & = & \frac{r^{2}-2Mr+a^{2}}{\rho^{2}}\left( dt-a \sin^{2}\theta d\phi \right)^{2}-\frac{\rho^{2}}{r^{2}-2Mr+a^{2}}dr^{2}\nonumber\\
& & -\rho^{2}d\theta^{2}-\frac{(r^{2}+a^{2})^{2}\sin^{2}\theta}{\rho^{2}}\left( d\phi-\frac{a}{r^{2}+a^{2}}dt \right)^{2}
\end{eqnarray}
\section{Concluding Remarks}
We have shown that it is possible at least in principle to derive the Kerr solution in the form~(\ref{eqn:niceKerr}) following some intuitive, physical arguments. Starting from an ellipsoidally symmetric flat space-time~(\ref{eqn:flatellip}), we used transformation~(\ref{eqn:transells}) from coordinates $(t,\phi,r,\theta)$ to coordinates $(T,\Phi,r,\theta)$ with the same dimension, along with an argument of reciprocity of two of the metric tensor components~(\ref{eqn:reciprocal}) to get the flat metric~(\ref{eqn:niceell}) in which the cross term $dt d\phi$ can arise if new metric tensor components $G'_{00}$ or $G'_{33}$ (or both) are changed in the presence of mass. We then used this metric to propose an ansatz from which we searched for the solution to the Einstein equations in an orthogonal metric form that resembles Schwarzschild solution~(\ref{eqn:schmet}). We have done this by demanding that time-time and radial-radial metric tensor components obey the same reciprocity relation~(\ref{eqn:reciprocal}) as Schwarzschild metric, and at the same time that the metric manifestly possesses the ellipsoidal symmetry of flat ellipsoidal space-time~(\ref{eqn:niceell}). It is interesting that the Kerr metric~(\ref{eqn:metricBL}) can be rewritten exactly in these coordinates, which results in orthogonal form of the Kerr metric. The derivation of orthogonal Kerr metric form~(\ref{eqn:niceKerr}) could be of pedagogical use, because this derivation is more intuitive than textbook derivations. Also, some of its features are more obvious in the form~(\ref{eqn:niceKerr}). For example, setting the outer event horizon\cite{Town97}
\begin{equation}
r_{+}=M+\sqrt{M^{2}-a^{2}}
\end{equation}
in metric~(\ref{eqn:niceKerr}), the Black Hole angular velocity\cite{Town97} becomes then manifest in the $d\Phi$ coordinate:
\begin{equation}
\Omega_{H}=\frac{a}{r^{2}_{+}+a^{2}}
\end{equation}
The motivation for this derivation emerged from the comparison of the metric tensor components of Schwarzschild solution~(\ref{eqn:schmet}), flat spherically symmetric space-time~(\ref{eqn:flatsph}), flat ellipsoidal space-time in the form~(\ref{eqn:niceell}) and Kerr metric in the form~(\ref{eqn:niceKerr}), which is presented here in the table~(\ref{tab:tabcomp}) below.
When one compares the corresponding metric tensor components, the most striking feature is the pattern of differences (and similarities) between the metric tensor components of different space-times within the columns of table~(\ref{tab:tabcomp}).
It would be interesting to check if full analogy is present between the ansatz~(\ref{eqn:flatsphans}) and~(\ref{eqn:kerransatz}). Does one get the same result that $e^{2\nu(r,\theta)}(-e^{2\sigma(r,\theta)})=-1$ for this metric while solving the Einstein equations? If this is the case, then coordinates $(T,\Phi,r,\theta)$ have a special role in rotating space-times. For the ellipsoidal space-time, these coordinates naturally give the possibility for rotation when one considers a rotating body with such a symmetry (ellipsoid of revolution). This connection between the reciprocity of time-time and radial-radial metric tensor components, rotation and orthogonality of the metric could maybe be an important one for some other axially-symmetric space-times too.
\begin{table}
\caption{Comparison of metric tensor components of several metric forms used in paper. In brackets are given the coordinates to which the new metric tensor components of the Kerr metric correspond.}
\label{tab:tabcomp}
\begin{tabular}[b]{lcccc}
& $dt^{2}(dT^{2})$ & $dr^{2}$ & $d\theta^{2}$ & $d\phi^{2}(d\Phi^{2})$ \\
flat spherical & $1$ & $-1$ & $-r^{2}$ & $-r^{2}\sin^{2}\theta$\\
Schwarzschild & $\displaystyle \frac{r^{2}-2Mr}{r^{2}}$ & $-\frac{r^{2}}{r^{2}-2Mr}$ & $-r^{2}$ & $-r^{2}\sin^{2}\theta$\\
flat ellipsoidal & $\displaystyle \frac{r^{2}+a^{2}}{r^{2}+a^{2}\cos^{2}\theta}$ & $\displaystyle -\frac{r^{2}+a^{2}\cos^{2}\theta}{r^{2}+a^{2}}$ & $\displaystyle -(r^{2}+a^{2}\cos^{2}\theta)$ & $\displaystyle -\frac{(r^{2}+a^{2})^{2}\sin^{2}\theta}{r^{2}+a^{2}\cos^{2}\theta}$\\
Kerr & $\displaystyle \frac{r^{2}-2Mr+a^{2}}{r^{2}+a^{2}\cos^{2}\theta}$ & $\displaystyle -\frac{r^{2}+a^{2}\cos^{2}\theta}{r^{2}-2Mr+a^{2}}$ & $\displaystyle -(r^{2}+a^{2}\cos^{2}\theta)$ & $\displaystyle -\frac{(r^{2}+a^{2})^{2}\sin^{2}\theta}{r^{2}+a^{2}\cos^{2}\theta}$
\end{tabular}
\end{table}

\begin{acknowledgments}
We thank Dr. Darko Kapor from the University of Novi Sad, Serbia, for reading the manuscript and giving suggestions regarding the english language. We are also grateful to Dr. Stanley Deser, Dr. Naresh Dadhich and Dr. Istv\'{a}n R\'{a}cz  for giving us very useful comments and pointing to some key references. This work was supported by the Serbian Ministry of Education and Science: Grant No 171009.
\end{acknowledgments}

\end{document}